# Large scale synthesis of 2D graphene oxide by mechanical milling of 3D carbon nanoparticles in air

Sandip Das[1], Subhamay Pramanik[1,#], Sumit Mukherjee[1], Tatan Ghosh[1], Rajib Nath[1], and, Probodh K. Kuiri[1,*]

[1]Department of Physics, Sidho-Kanho-Birsha University, Purulia – 723104, West Bengal, India

Abstract

Graphene oxide (GO) is one of the important functional materials. Large-scale synthesis of it is very challenging. Following a simple cost-effective route, large-scale GO was produced by mechanical (ball) milling, in air, of carbon nanoparticles (CNPs) present in carbon soot in the present study. The thickness of the GO layer was seen to decrease with an increase in milling time. Ball milling provided the required energy to acquire the in-plane graphitic order in the CNPs reducing the disorders in it. As the surface area of the layered structure became more and more with the increase in milling time, more and more oxygen of air got attached to the carbon in graphene leading to the formation of GO. An increase in the time of the ball mill up to 5 hours leads to a significant increase in the content of GO. Thus ball milling can be useful to produce large-scale two-dimensional GO for a short time.

\-\-\-\-\-\-\-\-\-\-\-\-\-\-\-\-\-\-\-\-\-\-\-\-\-\-
*Corresponding author, Electronic mail: probodh@skbu.ac.in
#Present address: Department of Physics, IIT Kharagpur, Kharagpur – 721302, India

# 1. Introduction

Graphene is very popular in the material science industry because of its large specific surface area, high intrinsic mobility [1,2], high Young's modulus [3], thermal conductivity [4], etc. Its high optical transmittance and good electrical conductivity make it a worthy candidate in the applications of transparent conductive electrodes [5,6] and many others. A variety of studies have been carried out which involve the use of chemically modified graphene in search of making new materials [7–10]. Graphene oxide (GO) is a key derivative of graphene, which exhibits outstanding mechanical, electrical, optical, thermal, and chemical properties due to these properties graphene finds many applications in electronics, sensors, solar cells, flexible displays, etc [11-14]. However, the performance of devices made of GO entirely depends on the quality of GO. This is because the structure and properties of GO are highly influenced by the synthesis method [15]. GO also have many direct applications, like it is used as polymer composites [7], antibacterial coatings [16,17], corrosion resistance coatings [18,19], energy-related materials, and paper-like materials [20–23], etc. Hence, the preparation of GO as well as the modification of its surface has attracted a lot of interest recently [24].

In 1859, Brodie has reported the synthesis of GO by reaction, in fuming $HNO_3$, of graphite and $KClO_3$ at 60 ˚C for 4 days [25]. Subsequently, epitaxial growth, thermal deoxidation, chemical vapour deposition, vertical cutting of carbon nanotubes, etc have been reported to produce GO [26,27]. Typically folding, cracking, and buckled layers are observed in the GO for many of the synthesized materials using usual techniques [28]. Also, each of the methods has its limitations such as low yield, prolonged process, and high cost. Additionally, the existing processes of preparation of GO are not free from impurities coming from the source of chemicals and solutions used [28]. Also, hazardous reagents during the oxidation of graphene can form [29]. Finally, the process of production of GO requires a tedious multistep usually. All these can lead to serious problems in structural integrity and hence the transfer of electrons in GO. Thus the big challenge is to produce large-scale GO with good quality. In order to achieve the production of GO on a commercial scale, the process needs to be simplistic, cost-effective, and environmentally sustainable. Recently, efforts have been made to develop more effective methods for the preparation of GO [28,30]. For example, Dash et al have reported the formation of GO by dry ball milling of graphite powder [28]. In their study, the formation of GO was found to require a time of 16 h. Fu et al have also reported the production of GO from graphite oxide using ball milling for a duration of 30 h [30]. Thus insights into the possibilities of preparation of large scale of GO have been reported employing the ball milling technique. It is to mention that in all the above-mentioned methods, graphene or graphite/graphite oxide is needed as precursors to produce GO. In the present study, we report the large-scale synthesis of GO of the thickness of a few nm from the carbon nanoparticles (CNPs) in carbon soot by ball milling in the air for 5 h of duration. The formation of two-dimensional (2D) GO was confirmed through high-resolution transmission electron microscopy (HRTEM), Raman spectroscopy, photoluminescence (PL) spectroscopy, and X-ray photoelectron spectroscopy (XPS). In the process of ball milling, oxygen from the air got attached to the carbon atom of graphene formed from the three-dimensional (3D) CNPs resulting in the formation of the 2D GO. This study can open up a new possibility for the production of large-scale 2D GO in a short duration. This is because ball milling is a very simple and effective method for the synthesis of various nanocrystalline and layered materials [31-34]. This method does not require any temperature rise externally for the production of required materials in grams of quantity.

## 2. Experiments

CNPs in carbon soots were prepared by firing a kerosene lamp and collecting the carbon soot on aluminum foil kept on top of the lamp wick when the lamp flame was coming out. Prior to firing the lamp, the kerosene lamp and the lamp wick were cleaned carefully using soap solution, de-ionized water, acetone, and finally de-ionized water. The black carbon deposited on the aluminum foils was collected and cleaned using de-ionized water followed by drying in a hot air oven. This way prepared carbon soot was then ball milled using Retsch PM 100 planetary ball mill at a speed of 300 rpm for time durations of 1 h, 3 h, and 5 h. For each case, a 1-minute interval was given after 10 minutes of continuous milling. This is to avoid sample heating due to the ball milling of the carbon soot. It is to mention that the reversal mode was used for two consecutive rotations. The carbon soot was kept inside a $ZrO_2$ bowl for milling. Also, the $ZrO_2$ ball of the diameter of 10 mm was used with the ball-to-sample mass ratio of 30:1. Hereafter the samples prepared this way will be referred to as $C_0$, $C_1$, $C_3$, and $C_5$ for before and after ball milling for milling times of 1 h, 3 h, and 5 h, respectively. The consideration of a maximum time (5 h) of milling is not the limit for the present study. Rather it is the initial study for the possible synthesis of 2D GO with a large amount using a simple procedure.

Transmission electron microscopy (TEM) and HRTEM images were recorded using Talos F200X G2 (Thermo Scientific) instrument to observe the structural and morphological changes before and after ball milling the carbon soot operating at 200 kV. The Raman spectroscopy for all of the samples was carried out using RENISHAW inVia Raman microscope with an inbuilt green laser ($\lambda$ = 532 nm) in the wavenumber range of 1100 to 3200 cm$^{-1}$. 5 mW of laser power was used in this experiment. PL measurements were carried out using the CARY Eclipse (Agilent) spectrofluorophotometer in the wavelength range of 350 nm to 550 nm with an excitation of 320 nm using a pulsed Xe lamp. XPS measurements were carried out using PHI 5000 Versa probe II with Al-K$_\alpha$ source (energy 1486.6 eV) in the range of 0 to 1100 eV. All the measurements were carried out at room temperature.

## 3. Results and Discussion

The TEM micrographs of the CNPs in soot, 1 h, 3 h, and 5 h ball-milled samples are shown in figures 1 (a), (b), (e), and (h), respectively, using the same image scale. The HRTEM image taken for the sample ball milled for a duration of 5 h is shown in the inset of figure 1(h). For the as-prepared carbon soot [figure 1(a)], CNPs of mainly two averages viz ~45nm and ~18 nm with nearly spherical shapes are found. The histogram of particle size distributions corresponding to the un-milled sample is shown in the inset of figure 1(a). After ball milling, these CNPs got transformed into layered structures. For clarity, the magnified views of a particular portion of each of the images of the milled samples are shown in figure 1(c), (f), and (i), respectively. The corresponding line profile is also shown in figure 1(d), (g), and (j), respectively. In the line profiles, the distance between two adjacent peaks (marked as parallel red lines) for the three milled samples (i.e. 1 h, 3 h, and 5 h milled samples) are about 2 nm, 1.5 nm, and 0.8 nm, respectively. As the thickness of a single layer of GO is ~0.7-0.8 nm [35,36], it is obvious that after milling, the samples contain stacks of a few layer structures. It is also easy to observe that the number of layers per stack in the sample is getting reduced as the duration of milling is getting increased. For the 1 h milled sample there are ~3 layers stacked together, for the 3 h milled sample, it reduces to ~2 layers, and for the 5 h milled sample, it is further reduced to ~1 layer of GO. These observations confirm that ball milling of carbon soot can produce graphitic structures, and further milling can reduce the number of layers, which can transform the graphitic structures into a few layered graphene-like structures with a possibility of formation of GO. It is mentioned that the ball milling of carbon nanoparticles prepared by kerosene soot also produced

GO [37]. Thus the present study is consistent with the earlier reported data. The schematic diagram for the synthesis of GO by ball milling of carbon nanoparticles in carbon soot is shown in figure 2. During ball milling in air ambient, the amorphous carbon nanoparticles get transformed into multi-layered structures of polyaromatic carbon [38].

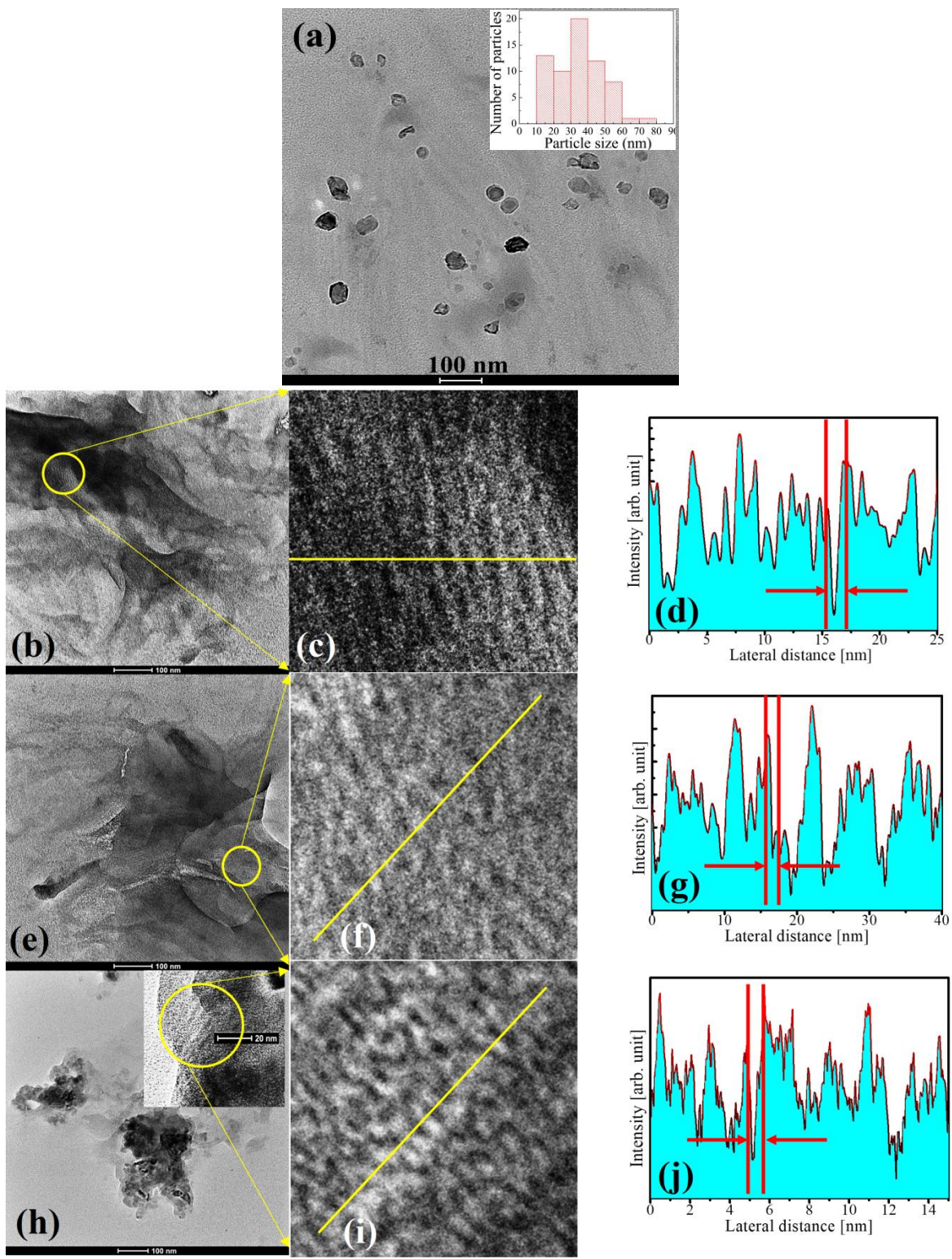

Figure 1. TEM image of (a) as prepared carbon soot, (b) 1 h ball milled sample, (e) 3 h ball milled sample, (h) 5 h ball milled sample. HRTEM of 5 h ball milled sample is shown in the inset of figure 5(h). The magnified view of (c) 1 h ball milled sample, (f) 3 h ball milled sample, (i) 5 h ball milled sample. The line profile of (d) 1 h ball milled sample, (g) 3 h ball milled sample, (j) 5 h ball milled sample.

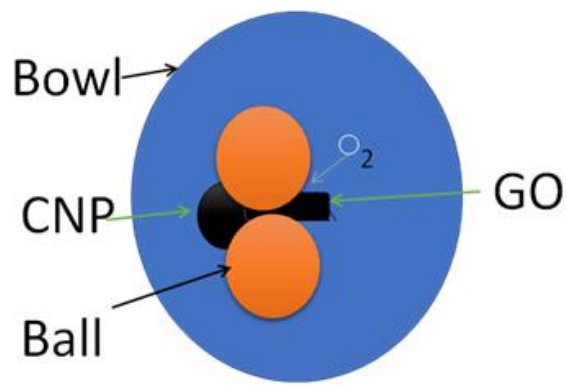

Figure 2 : Schematic diagram of synthesis of GO using ball milling of carbon nanoparticles.

Raman spectroscopy is a very sensitive tool to the structural changes of solids [39,40]. Especially for carbon, it is possible to know the presence of types of carbon viz amorphous carbon, six-fold rings, in-plane graphitic order, diamond-like carbon, etc [41]. In fact, the changes in these structures can also be detected using Raman spectroscopy efficiently. It is to mention that the carbon soot containing CNPs was prepared using the burning of kerosene followed by ball milling for different durations. It is expected that these samples can have different carbon structures. We have tried to know in detail the type of carbon present in the soot and their modification by ball milling using Raman spectroscopy also. Figure 3(a) shows the Raman spectra of the CNPs in carbon soot within the frequency range of 1000 $cm^{-1}$ to 3200 $cm^{-1}$. Two distinct Raman bands with peaks at about 1347 $cm^{-1}$ and 1600 $cm^{-1}$ are seen in the carbon soot before ball milling. The first Raman peak is relatively broader than the second peak. These two peaks in the Raman spectrum are well known as the *D* and *G* bands, respectively [42]. For the carbon soot after ball milling for time durations of 1 h, 3 h, and 5 h, these two bands are also seen to appear in the Raman spectra. However, the structure (peak position, FWHM, peak intensity, etc) of these bands changes with an increase in the time of milling.

The appearance of *D* and *G* bands in the Raman spectra indicate clearly that carbon particles were produced in the process [42]. It is well known that the *G* band corresponds to the graphite structures with $sp^2$ bonding and the ordered *G* band possesses the $E_{2g}$ phonon at the Brillouin zone centre [43]. The in-plane vibration of graphite structure is identified as the *G* band in the Raman spectrum. These vibrations of carbon atoms are shown schematically in Figure 4(a). The $E_{2g}$ mode of vibration does not always require the six-fold ring structure of graphite [as shown in figure 4(b)]. In this mode, pairs of $sp^2$ carbon atoms undergo the in-plane bond-stretching motion and can always be found in the range of 1500–1630 cm$^{-1}$ [44]. On the other hand, the *D* band arises due to the breathing modes of six-atom rings ($A_{1g}$ symmetry) and requires a defect for its activation [41]. The *D* band comes from the transverse optical phonons around the ***K*** and ***K$^{/}$*** points in the first Brillouin zone. The vibration of the carbon atoms in the six-fold ring structure of graphite corresponding to the *D* band is shown schematically in figure 4(b). The *D* band can vary with the photon excitation energy and can be dispersive due to a Kohn anomaly at ***K*** [45-48]. In the present study, the excitation of the samples was fixed (532 nm) and we can make a fair comparison of the Raman spectra for different samples, especially for the *D* band. Thus from the above discussions, it is confirmed that all the samples (carbon soot before and after ball milling) contain carbon in the form of in-plane graphite ordering with a lot of disorders (or six-fold rings in amorphous carbon) in it [49,50].

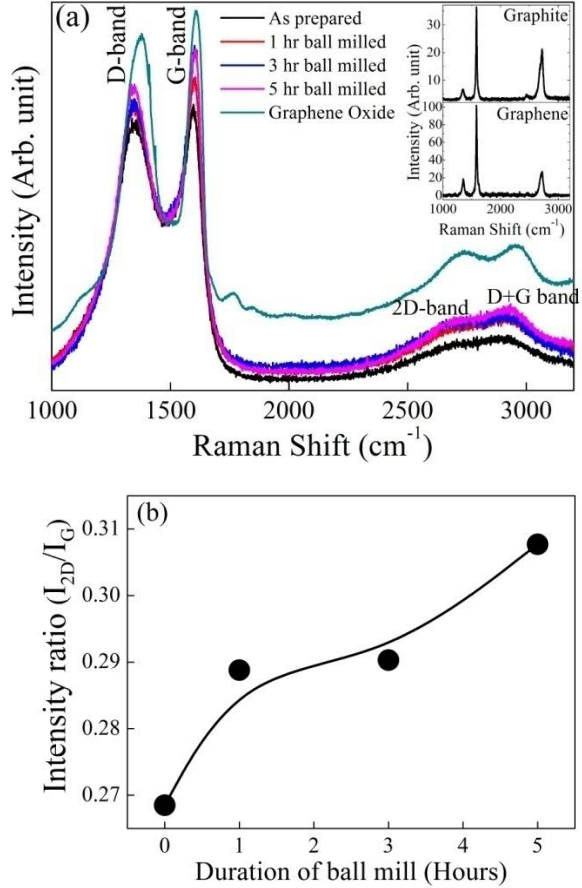

Figure 3. (a) Raman spectra of as prepared and ball milled samples. For comparison, the Raman spectrum of graphene oxide is also plotted. Inset shows the Raman spectra for graphene and graphite. (b) Variation of intensity ratio ($I_{2D}/I_G$) with duration of ball milling. Solid line is to guide the eye.

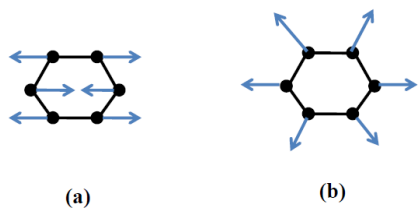

Figure 4: Schematic representations of (a) the $E_{2g}$ mode of vibration of *G* band and (b) the $A_{1g}$ breathing mode of variation of *D* band.

For comparison, the Raman spectrum of GO is also plotted in figure 3(a). The observed Raman spectra of the carbon samples in the present study match well with that of GO. Further, the Raman spectra for graphite and graphene are also shown as insets in figure 3(a). The observed *D* and *G* bands are also seen in the graphite and graphene. Based on *D* and *G* bands it is not possible to conclude about the formation of GO after ball milling. For this purpose, we need to study the Raman spectra in the frequency region of 2000 cm$^{-1}$ to 3200 cm$^{-1}$. A broad hump ranging in the frequency region of ~2400 cm$^{-1}$ and beyond in the Raman spectrum is found for the CNPs in carbon soot before ball milling. This hump is seen to develop into two bands at ~2689 cm$^{-1}$ and ~2925 cm$^{-1}$ after ball milling. Two clear bands are seen for the sample after ball milling of 5 h and these bands match with the bands of GO [figure 3(a)]. The appearance of the Raman band with a peak at ~2689 cm$^{-1}$ represents the graphene-like structure and is attributed due to the double resonance transitions (i.e. 2*D* mode) which results from the production of two phonons having opposite momenta [51]. The other band centered at ~2925 cm$^{-1}$ is the defect-activated *D+G* mode [52]. These observations indicate that GO is produced by ball-milling the CNPs in carbon soot. To study the effects of ball milling on the production of GO, we have plotted the ratio of the intensity of the 2*D* peak and the *G* peak in figure 3(b). The intensity ratio increases with an increase in the time of ball milling indicating that more and more graphitic structures are getting converted to GO. These results are consistent with the XRD [37] and TEM studies. These observations confirm that GO can be produced via ball milling of carbon soot produced by kerosene combustion method, which is one of the easiest methods of producing GO. We believe that the quantity and quality of the produced GO can be improved by controlling the various parameters during the process of milling.

For a clear understanding of the formation of the layered structure and GO by ball milling the CNPs in carbon soot, we have fitted the *D* and *G* bands of Raman spectra using Lorentz plus

Breit-Wigner-Fano (BWF) lines [53]. Usually, the Raman spectra are fitted using the Lorentzian line shapes when there is no asymmetry in the experimental spectra [54-56]. However, in case, the asymmetry exists in the Raman spectrum, the fitting is done using BWF lines [54-56]. As the G band tails toward the lower frequency, the asymmetry arises which is not found in the D band (which is symmetric). To take care of this asymmetry, the BWF function was found far better than using Lorentzian (or Gaussian) line when fitting the Raman spectrum for the G band. The BWF line is given by [53]

$$I(\omega) = \frac{I_0 \left\{1 + \frac{2(\omega - \omega_0)}{Q\Gamma}\right\}^2}{1 + \left\{\frac{2(\omega - \omega_0)}{\Gamma}\right\}^2} \qquad (1)$$

where $I(\omega)$ is the frequency-dependent intensity, $I_0$ is the peak intensity, $\omega_o$ is the peak position, $\Gamma$ is the full width at half maximum (FWHM) of the peak, and $Q^{-1}$ is the BWF coupling coefficient. In the limit $Q^{-1} \to 0$, the BWF line becomes the Lorentz line. The asymmetric shape of the BWF line can be explained as the coupling of a discrete mode to a continuum [57]. The -$\Gamma Q$ is the coupling parameter and is a measure of in-plane graphitic order. The limit -$\Gamma Q \to 0$ is defined as the point where in-plane graphitic ordering is complete and three-dimensional ordering begins [58]. From the literature, it is found that the pair of BWF plus Lorentz line fits the Raman spectra of all types of carbons in a very excellent way [53,59]. The experimental Raman spectra along with the Lorentz function and BWF function fittings are shown in figure 5(a) – (d) for the samples $C_0$, $C_1$, $C_3$, and $C_5$, respectively. All the spectra are resembled well by these lines. Below we discuss the different fitting parameters.

Figure 6(a) shows the variations of the peak positions of the D and G bands as a function of milling time. The peak position of the D band is found to be nearly constant at different times. However, the peak position of the G band ($\omega_o$) increases with an increase in time, i.e. the peak shifts towards the higher frequency side. Such a shift of the G band indicates vibrations of more carbon atoms in the in-plane as a result of the graphitization of carbon soot with ball milling time. The Inset of Figure 6(a) shows the variations of $\omega_o$ and $\omega_{max}$ with the time of milling. The $\omega_{max}$ is given by [53]

$$\omega_{max} = \omega_0 + \frac{\Gamma}{2Q} \qquad (2)$$

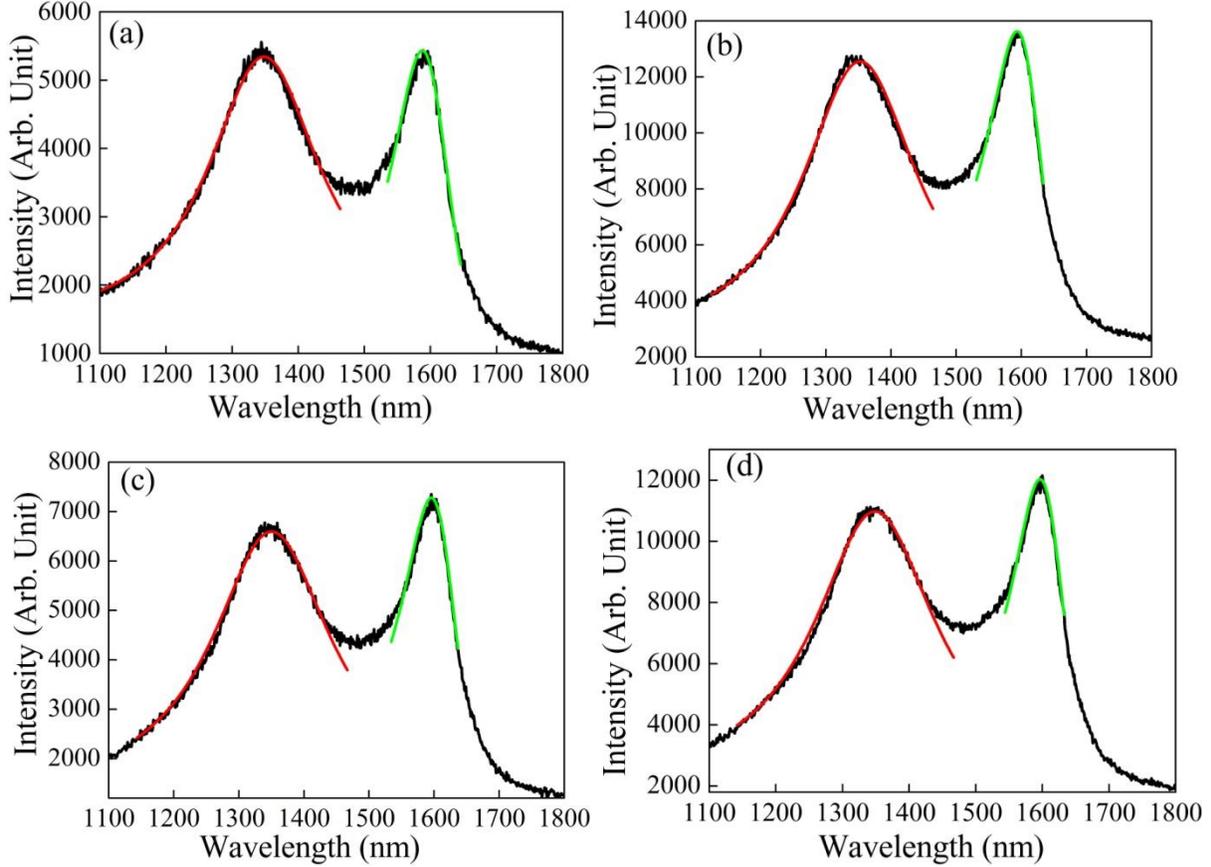

Figure 5. Experimental Raman spectra (black line in each figure) of samples $C_0$ (a), $C_1$ (b), $C_3$ (c), and $C_5$ (d) along with their fittings using a Lorentzian function for *D* band (red line in each figure) and a BWF function [eq, (1)] for *G* band (green line in each figure) corresponding to each of the samples.

This relation lowers the $\omega_o$ to $\omega_{max}$ (as *Q* is negative) which is the frequency at which the *G* band is maximum. Observation of a higher value of $\omega_o$ than the apparent peak maximum $\omega_{max}$ is because $\omega_o$ is the undamped mode [57]. As seen from the graph, the $\omega_{max}$ appears about 12-15 cm$^{-1}$ below the $\omega_o$.

Figure 6(b) shows the variations of the width (FWHM, $\Gamma$) for both *D* and *G* bands as a function of the time of milling. The $\Gamma$ decreases nearly linearly for the *G* band whereas it increases exponentially for the *D* band with the time of milling. A decrease in the width of the *G* band is indicative of the graphitization of the carbon soot [54]. And an increase in the width of the *D* band with time is indicative of a reduction in the number of six-fold rings in CNPs. The peak positions and FWHM of the *D* and *G* bands as a function of the time of milling is also provided in table 1 of supplementary file. In figure 6(c), the variation of the coupling parameter ($-\Gamma/Q$) of the BWF line is shown. An increase in ball milling time leads to a decrease in the value of the coupling parameter linearly. As mentioned above, the coupling parameter is the measure of the in-plane ordering of the graphite – the lower the value of the coupling parameter higher the in-plane graphitic order. A linear decrease of the value of $-\Gamma/Q$ indicates that after ball milling the carbon soot, in-plane graphitic order increases. Furthermore, the in-plane graphitic order increases with an increase in the time of ball milling. Such in-plane graphitic order can also be identified from the variation of $I_D/I_G$ (intensity ratio of maxima of the $D$ band to that of the $G$ band) as a function of time of ball milling. The variation of $I_D/I_G$ with the time of ball milling is shown in figure 6(d).

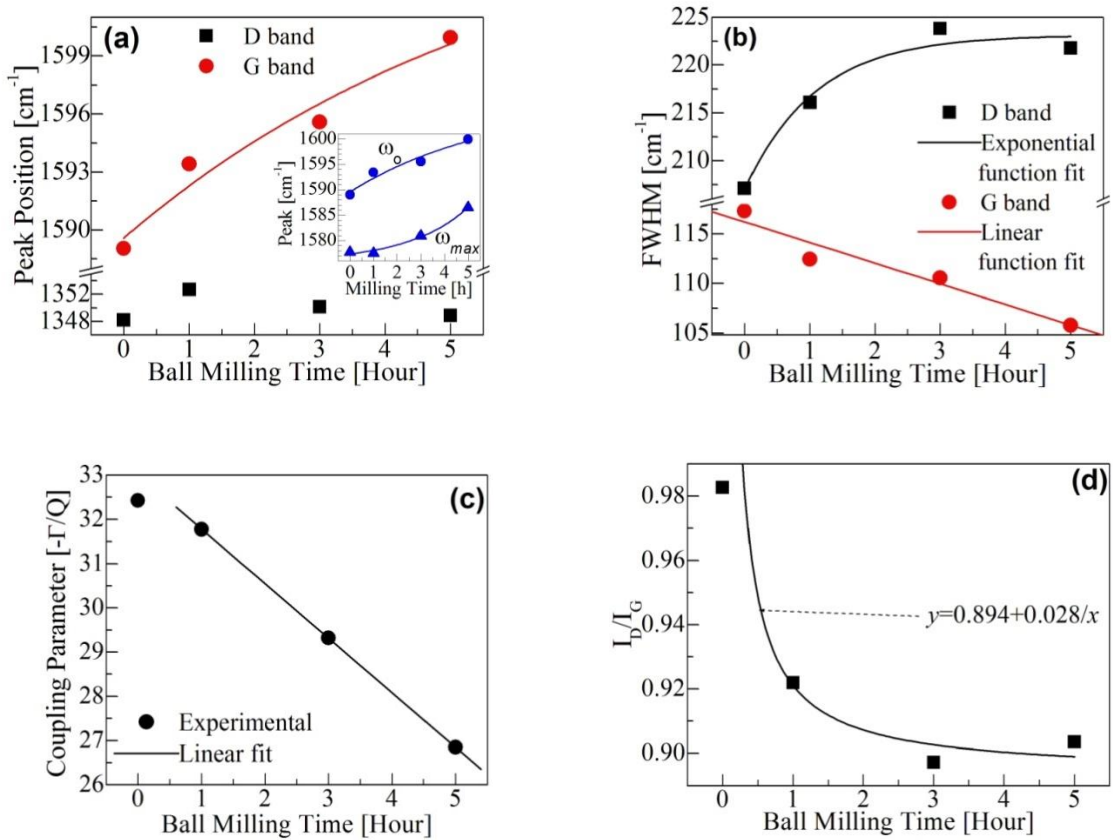

Figure 6. Variations of Raman (a) peak positions of $D$ and $G$ bands, (b) FWHM of $D$ and $G$ bands, (c) Coupling parameter ($-\Gamma/Q$) of $G$ band, and (d) the ratio of peak intensities of $D$ and $G$ bands as a function of ball milling time. Solid curves in (a) are the exponential function fits of the corresponding experimental data points. Inset of (a) shows the variations of $\omega_o$ and $\omega_{max}$ as a function of milling time.

$I_D/I_G$ decreases very fast initially and then shows a saturation behavior. To guide the variation of the experimental data points with a time of milling, $t$, we have drawn a curve as $I_D/I_G \sim 1/t$ in the same figure. This line indicates that $I_D/I_G$ follows $1/t$ functional variation. This variation is similar to the variation [60]

$$I_D/I_G = C(\lambda)/L_a \qquad (3)$$

where $C$ ($\approx$4.4 nm for $\lambda$=415.5 nm) and $L_a$ are the constant and the size of the graphite crystallites in a sample, respectively. This is TK equation (Tuinstra and Koenig). Hence $t$ becomes equivalent to $L_a$ in the present study. According to eq (3), the $I_D/I_G$ will decrease with an increase in $L_a$. A sample containing both in-plane graphitic order and amorphous carbon with six-fold rings will show two Raman bands viz $D$ and $G$ bands. If there is any change in the in-plane graphitic order or six-fold rings, then corresponding changes in the intensity of $G$ or $D$ bands will occur following the TK relation. A reduction of the number of six-fold rings (disorders) or an increase in the in-plane graphitic order in the sample will lead to a corresponding decrease in the intensity of the $D$ band. From the observed variation of $I_D/I_G$ with $t$ in the present study, it appears that the disordering decreases and/or in-plane graphitic order increases.

It is also important to mention that graphite structures are much more stable against mechanical impact. At this stage, it is important to understand how in-plane graphitic order increases in the carbon particle with an increase in ball milling time. In this case, the increase of the in-plane graphitic order cannot be explained by the self-assembly of small fragments of carbon particles in carbon soot. Zhang et al have shown that the curvature of the transition regions of the closed-shell structures was enlarged and the flat graphite sheets became curved during the ball-milling of carbon black [61]. Therefore, the polyhedral structure tends to form a spherical one. It is also known that from an energetic point of view, carbon *onions* (which are spherical in structure) are much more stable than other forms of CNPs [62,63]. The high energy mechanical impact during ball milling can induce the transformation of the unstable polyhedral structure to a more stable spherical structure spontaneously driven by the minimization of energy of the carbon system. In view of this, Zhang et al have proposed a transformation model to explain how the ball-milling induced the enhancement of the graphite crystal degree in the closed-shell structure [61]. Ball milling provides the required energy to acquire the in-plane graphitic order in the CNPs reducing the disorders in it. We also believe that a similar situation of the Zhang et al is true for the present CNPs in carbon soot and their ball milling for different times. More and more in-plane graphitic orders were formed with a corresponding reduction of the six-fold rings in the CNPs with an increase in the time of milling. Such changes in the CNPs resulted in the observed changes in the *G* and *D* bands of the Raman spectra. Moreover, the presence of oxygen in air leads to the formation of GO by ball-milling the CNPs in carbon soot.

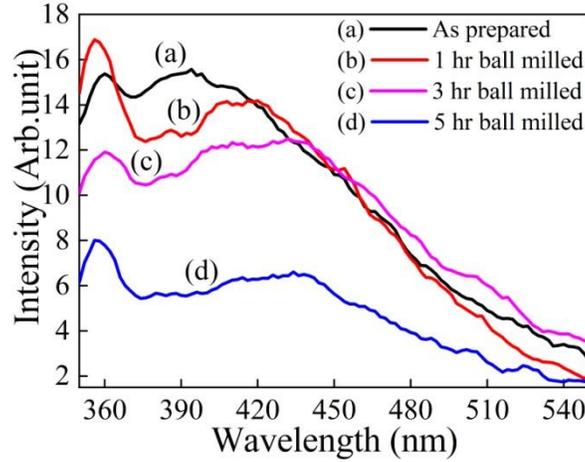

Figure 7. PL emission spectra of samples excited at 320 nm.

Figure 7 shows the room temperature PL emission spectra of all the samples excited at 320 nm. Two PL emissions bands at ~360 nm (Peak 1) and ~394 nm (Peak 2) are seen for CNPs in carbon soot before ball milling. The first emission band belongs to the ultra-violet (UV) region which is relatively narrow. The second emission band which is a broad one belongs to the violet region. Similar two PL emission bands are also observed for other samples after ball milling for different milling times viz. 1 h, 3 h, and 5 h. Two PL emissions from CNPs synthesized using laser ablation were also observed experimentally by Nguyen et al. The PL emissions from these CNPs were controlled by subsequent UV irradiation in an oxygen atmosphere [64]. The first peak for all the samples is almost fixed at about 358 nm (this is shown in figure 8 as black-colored filled squares), while the second peak is seen to shift towards a higher wavelength side (416 nm, 432 nm, and 434 nm for the carbon particles prepared by ball milling for time durations of 1 h, 3 h, and 5 h, respectively) with an increase in ball milling time. This shift is about 40 nm (from 394 nm to 434 nm) and is redshifted. The variations of the PL peaks with the time of ball milling are shown in figure 8 (red-colored filled circles). This variation is well-fitted using an exponential function and is also shown as the (red) solid line in the same figure.

Observation of two PL emissions at different positions indicates that the origin of such PL is due to two independent emission centres in CNPs. Though the exact origin of the PL emissions from CNPs is not known, the generally accepted reasons for such PL emissions are different surface states, which are caused by different functional groups [64,65]. Based on the observation of the similar nature of two peak structures in the present study and that of Nguyen et al [64], we also believe that such PL emissions originated from abundant surface functional groups on the surface of carbon particles rather than its carbogenic core. With a detailed study,

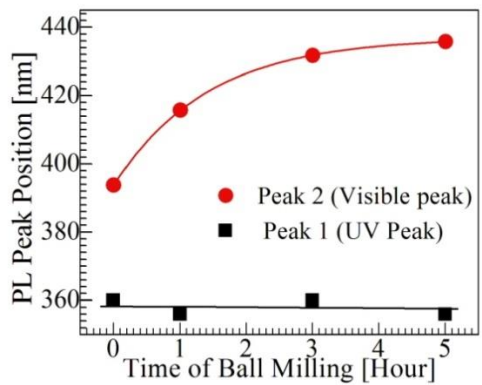

Figure 8. Variations of PL emission peaks with time of ball milling. Solid lines are linear and exponential function fits for UV peak (Peak 1) and visible peak (Peak 2), respectively.

Nguyen et al [64] have shown that the origin of the UV PL peak (Peak 1) is due to the C=O functional group and the origin of the visible PL peak (Peak 2) is due to C–O functional group forming two different surface states. We believe that the origin of UV PL emission in the present study is due to the C=O functional group and the origin of the visible PL peak is due to the C–O functional group. According to Nguyen et al, both of these emissions are independent of carbogenic-core size [64]. However, we have observed a redshift of the visible peak (from violet to red and then blue) with an increase in ball milling time while the UV PL emission peak position is independent of the ball milling time. It was also reported by A. P. Demchenko that the violet emission is due to the surface states of carbon particles formed as a result of nitrogen and oxygen-containing functional group and the blue emission is due to the formation of oxygen-containing functional groups [66]. As mentioned in the Experimental section, the carbon soot (containing CNPs) was prepared in an ambient atmosphere. This was followed by ball milling in the air. Thus there is the possibility of the formation of surface functional groups containing both nitrogen and oxygen during the synthesis processes. Due to the presence of these functional groups in carbon, two PL emissions were observed in the present study. With an increase in the time of ball milling, more and more functional groups containing oxygen (likely of type of C–O) formed by replacing nitrogen. This can result in a redshift of the visible peak. As mentioned above, the thickness of carbon particles was reduced (the formation of layered structure increases) with an increase in the time of ball milling. Thus it is likely to form more C–O with an increase in ball milling time as the surface area becomes more with a thinner layer of graphite which results in the formation of GO. The formation of more and more GO leads to a redshift of the visible PL emission peak.

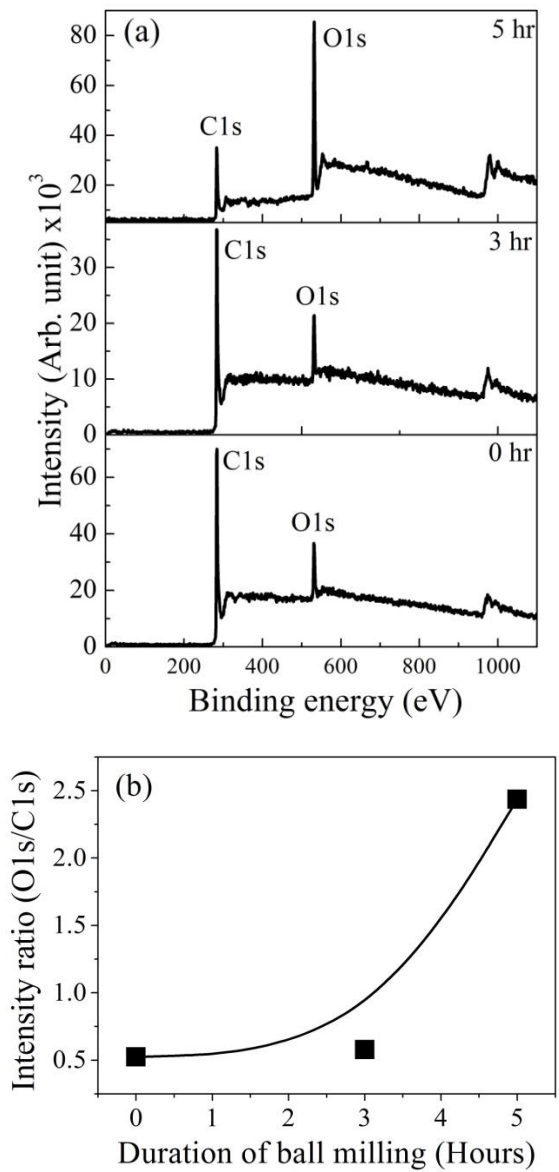

Figure 9: (a) XPS spectra of as-prepared carbon soot, ball milled for 3 h and 5 h. (b) Variation of ratio of intensity of O1s to that of C1s peaks. Solid line is to guide the eye.

The XPS is the most widely used technique for evaluation of the chemical states and the presence of various functional groups, in other words, to extract detailed information about the bonding structure, surface chemistry, and composition of surfaces and interfaces [67,68]. The chemical states of the carbon in the samples were investigated using XPS. Figure 9 (a) shows the XPS spectra of carbon soot before and after ball milling for a duration of 3 h and 5 h. All the spectra show two intense peaks at ~284 eV and ~532 eV with a small peak at ~975 eV. The XPS peaks at ~284 eV and ~532 eV are due to the C1s and O1s [69]. The presence of O is found even in the as-prepared carbon soot. As the carbon soot was prepared in the air, the oxygen get incorporated into carbon forming C-O and C=O bonds. After ball milling, the intensity corresponding to ball-milled samples increases significantly. Figure 9(b) shows the variation in the ratio of the intensity O1s peak to that of the C1s peak. This ratio is seen to increase significantly after ball-milling the carbon soot for a duration of 5 h. This confirms that a lot of oxygen from the atmosphere got attached to the carbon atoms of the graphene forming the GO during ball milling [70]. The XPS spectrum corresponding to sample $C_5$ is quite similar to that of the GO [68]. To further confirm the formation of GO, we have deconvoluted the C1s XPS peak using Gaussian line shapes. Figure 10(a) shows the Gaussian function fitted C1s peak. For sample $C_0$ (carbon soot before ball milling), three Gaussian peaks at ~284.5 eV, ~285.7 eV, and ~288.5 eV are found to reproduce the experimental data. The first peak at ~284.5 eV is due to the C-C $sp^2$ carbon bond [71]. The other two peaks are due to the C-O and C=O bonds in the samples [72]. Similar three Gaussian functions were also used to fit the C1s peak for sample $C_3$ (ball milled for 3 h). However, for sample $C_5$ (ball milled for 5 h), four Gaussian functions were used to fit the experimental data. The addition peak at ~286.2 eV is due to the formation of the C-OH functional group in carbon [71]. This bond is formed in the sample due to the attachment of some more oxygen functional groups to the carbon atoms during ball milling [73].

The intensity of the deconvoluted peak corresponding to the C-C bond is seen to decrease with an increase in ball-milling time. However, the intensities of deconvoluted peaks corresponding to C-O and C=O bonds are seen to increase with an increase in ball milling time. The variations in the intensity of these peaks are shown in figure 10(b). The intensity of the C-C bond monotonously decreases. However, the same for C-O and C=O bonds increases with an increase in the time of ball milling. The continuous decrease of the C-C bond and increase in the intensity of the other bonds viz. C-O and C=O are indications of an increase in the oxidation level in the samples due to milling in air ambient [71]. These observations along with TEM, Raman spectroscopy, and PL spectroscopy studies unambiguously confirm the formation of GO in the ball-milled carbon soot, especially ball milling for 5 h. It is to mention that in this process (ball milling of carbon nanoparticles in the air) large-scale (amount) of GO can be synthesized.

Although 3 g of carbon soot containing carbon nanoparticles was taken for ball milling in the present study, the amount of the carbon nanoparticles can be taken larger than 3 g in this process producing GO. From figures 6, 8, 9(b), and 10(b), one can see that the complete oxidation of carbon atoms in the carbon nanoparticles was not taken place. This indicates that ball milling of more than 5 h can produce more GO in the samples. It will be interesting to repeat the synthesis process beyond the ball milling time of 5 h (which is the maximum time taken in the present study). We also plan to work in this direction in the future.

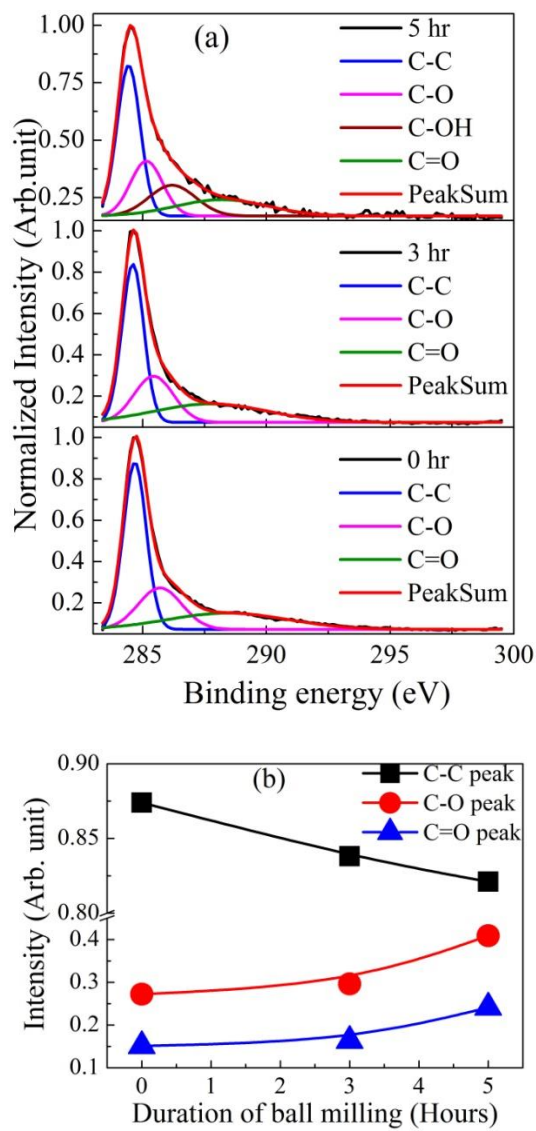

Figure 10: (a) XPS spectra of C1s band with Gaussian functions fits, (b) Variation of intensity of C-C peak, C-O peak, and C=O peak with duration of milling. Solid lines are to guide the eye.

## 4. Conclusions

Ball milling in the air ambient of CNPs present in carbon soot was found to produce 2D GO. The thickness of the GO layer was seen to decrease with an increase in milling time. Ball milling provided the required energy to acquire the in-plane graphitic order in the CNPs reducing the disorders in it. As the surface area of the layered structure became more and more with the increase in milling time, more and more carbon got attached to the oxygen present in the air leading to the formation of GO in the process of ball milling. The formation of CNPs and GO was confirmed through TEM, Raman, PL, and XPS spectroscopy measurements. An increase in the time of ball milling up to 5 h led to a significant increase in the content of GO in CNPs converting to the layered structure with reduced thickness. Thus ball milling can be a suitable method to produce large-scale GO which is otherwise a big challenge today to synthesize large-scale GO.

**Acknowledgments**

The authors would like to thank the central instrumental facility, Indian School Mines, Dhanbad and the central research facility, IIT Kharagpur, Kharagpur for extending their help for the TEM facility and XPS facility, respectively. We would also like to acknowledge the Department of Science and Technology, Govt. of India for partial financial support through FIST project (No.: SR/FST/PS-I/2020/159). SD acknowledges the Council of Scientific and Industrial Research, Govt of India viz file No.: 09/1156(0003)/2017-EMR-I.

**Conflict of Interest**

The authors have no conflicts of interest to declare that are relevant to the content of this article.

**Data Availability**

The authors declare that all the data supporting the findings of this study are available within the article.

# Supplementary data
for
## Large scale synthesis of 2D graphene oxide by mechanical milling of 3D carbon nanoparticles in air

Sandip Das[1], Subhamay Pramanik[1,#], Sumit Mukherjee[1], Tatan Ghosh[1], Rajib Nath[1], and, Probodh K. Kuiri[1,*]

[1]Department of Physics, Sidho-Kanho-Birsha University, Purulia – 723104, West Bengal, India

Table 1: Peak position and FWHM of the Raman spectra for D and G bands of carbon nanoparticles before and after ball milling.

|  | As prepared | | 1 h ball milled | | 3 h ball milled | | 5 h ball milled | |
| --- | --- | --- | --- | --- | --- | --- | --- | --- |
|  | D-band | G-band | D-band | G-band | D-band | G-band | D-band | G-band |
| Peak position (cm$^{-1}$) | 1348 | 1589.05 | 1352.7 | 1593.42 | 1350.2 | 1595.6 | 1348.8 | 1599.96 |
| FWHM (cm$^{-1}$) | 207 | 117.3 | 216 | 112.4 | 223.8 | 110.5 | 221.8 | 105.8 |